\documentstyle[12pt,amssymb]{article}
\begin{document}
\font\frak=eufm10 scaled\magstep1
\font\fak=eufm10 scaled\magstep2
\font\fk=eufm10 scaled\magstep3
\font\black=msbm10 scaled\magstep1
\font\bigblack=msbm10 scaled\magstep 2
\font\bbigblack=msbm10 scaled\magstep3
\font\scriptfrak=eufm10
\font\tenfrak=eufm10
\font\tenblack=msbm10


\def\biggoth #1{\hbox{{\fak #1}}}
\def\bbiggoth #1{\hbox{{\fk #1}}}
\def\sp #1{{{\cal #1}}}
\def\goth #1{\hbox{{\frak #1}}}
\def\scriptgoth #1{\hbox{{\scriptfrak #1}}}
\def\smallgoth #1{\hbox{{\tenfrak #1}}}
\def\smallfield #1{\hbox{{\tenblack #1}}}
\def\field #1{\hbox{{\black #1}}}
\def\bigfield #1{\hbox{{\bigblack #1}}}
\def\bbigfield #1{\hbox{{\bbigblack #1}}}
\def\Bbb #1{\hbox{{\black #1}}}
\def\v #1{\vert #1\vert}             
\def\ord#1{\vert #1\vert} 
\def\m #1 #2{(-1)^{{\v #1} {\v #2}}} 
\def\lie #1{{\sp L_{\!#1}}}                                             
\def\pd#1#2{\frac{\partial#1}{\partial#2}}
\def\pois#1#2{\{#1,#2\}}
\def\set#1{\{\,#1\,\}}             
\def\<#1>{\langle#1\rangle}        
\def\>#1{{\bf #1}}                
\def\f(#1,#2){\frac{#1}{#2}}
\def\cociente #1#2{\frac{#1}{#2}}
\def\braket#1#2{\langle#1\mathbin\vert#2\rangle} 
\def\brakt#1#2{\langle#1\mathbin,#2\rangle}                                     
\def\dd#1{\frac{\partial}{\partial#1}} 
\def\bra #1{{\langle #1 |}}
\def\ket #1{{| #1 \rangle }}
\def\ddt#1{\frac{d #1}{dt}}
\def\dt2#1{\frac{d^2 #1}{dt^2}}
\def\matriz#1#2{\left( \begin{array}{#1} #2 \end{array}\right) }
\def\Eq#1{{\begin{equation} #1 \end{equation}}}

\def\bw{{\bigwedge}}                                                                                                                            
\def\hut{{\scriptstyle \wedge}}                                 
\def\dg{{\goth g^*}}                                                                                                            
\def\Cdg{{C^\infty (\goth g^*)}}
\def\poi{\{\:,\}}                                                               
\def\qw{\hat\omega}                
\def\FL{{\sp F}L}                 
\def\hFL{\widehat{{\sp F}L}}      
\def\XHMw{\goth X_H(M,\omega)} 
\def\XLHMw{\goth X_{LH}(M,\omega)}                  
\def\ea{\varepsilon_a}
\def\ep{\varepsilon}
\def\mitad{\frac{1}{2}}
\def\x{\times}  
\def\cinf{C^\infty} 
\def\forms{\bigwedge}                 
\def\onda{\tilde}
\def\orb{{\sp O}}

\def\a{\alpha}
\def\g{{\gamma }}                  
\def\G{{\Gamma}}
\def\La{\Lambda}                   
\def\la{\lambda}                   
\def\w{\omega}                     
\def\W{{\Omega}}                   
\def\ltimes{\bowtie} 
             
\def\roc{{\tilde{\cal R}}}                                                                              
\def\cl{{\cal L}}                                                                                                                                                       
\def\V{{\sp V}}                                 
\def\F{{\sp F}}
\def\cv{{{\goth X}}}                    
\def\LG{\goth g}
\def\X{{{\goth X}}}                     
\def\R{{\hbox{{\field R}}}}             
\def\big R{{\hbox{{\bigfield R}}}}
\def\bbig R{{\hbox{{\bbigfield R}}}}
\def\C{{\hbox{{\field C}}}}         
\def\Z{{\hbox{{\field Z}}}}             
\def\N{{\hbox{{\field N}}}}         

\def\ima{\hbox{{\rm Im}}}                                                                                       
\def\dim{\hbox{{\rm dim}}}                                                                      
\def\End{\hbox{{\rm End}}} 
\def\Tr{\hbox{{\rm Tr}}} 
\def\tr{{\hbox{\rm\small{Tr}}}}                                                                                                 
\def\lin{{\hbox{Lin}}}
\def\vol{{\hbox{vol}}}  
\def\Hom{{\hbox{Hom}}}
\def\div{{\hbox{div}}} 
\def\rank{{\hbox{rank}}}
\def\Ad{{\hbox{Ad}}}
\def\ad{{\hbox{ad}}}
\def\CoAd{{\hbox{CoAd}}}
\def\coad{{\hbox{coad}}}                                                                
\def\Rea{\hbox{Re}}                             
\def\id{{\hbox{id}}}                    
\def\Id{{\hbox{Id}}}
\def\Int{{\hbox{Int}}}
\def\Ext{{\hbox{Ext}}}
\def\Aut{{\hbox{Aut}}}
\def\Card{{\hbox{Card}}}
\def\SODE{{\small{SODE }}}


\newtheorem{teor}{Teorema}[section]
\newtheorem{cor}{Corolario}[section]
\newtheorem{prop}{Proposici\'on}[section]
\newtheorem{definicion}{Definici\'on}[section]
\newtheorem{lema}{Lema}[section]

\newtheorem{theorem}{Theorem}
\newtheorem{corollary}{Corollary}
\newtheorem{proposition}{Proposition}
\newtheorem{definition}{Definition}
\newtheorem{lemma}{Lemma}
\def\be{\begin{equation}}
\def\ee{\end{equation}}
\def\Eq#1{{\begin{equation} #1 \end{equation}}}
\def\R{\Bbb R}
\def\C{\Bbb C}
\def\Z{\Bbb Z}
\def\d{\partial}

\def\la#1{\lambda_{#1}}
\def\teet#1#2{\theta [\eta _{#1}] (#2)}
\def\tede#1{\theta [\delta](#1)}
\def\N{{\frak N}}
\def\Wei{\wp}
\def\Hil{{\cal H}}

\font\frak=eufm10 scaled\magstep1

\def\bra#1{\langle#1|}
\def\ket#1{|#1\rangle}
\def\goth #1{\hbox{{\frak #1}}}
\def\<#1>{\langle#1\rangle}
\def\cotg{\mathop{\rm cotg}\nolimits}
\def\wt{\widetilde}
\def\const{\hbox{const}}
\def\grad{\mathop{\rm grad}\nolimits}
\def\Div{\mathop{\rm div}\nolimits}
\def\braket#1#2{\langle#1|#2\rangle}
\def\Erf{\mathop{\rm Erf}\nolimits}

\centerline{\Large \bf Non--hermitian oscillator--like Hamiltonians } 

\medskip

\centerline{\Large \bf  and $\lambda$-coherent states revisited}

\vskip 2cm

\centerline{\sc Jules Beckers$^{\ddag}$\footnote{email: Jules.Beckers@ulg.ac.be}, 
Jos\'e F. Cari\~nena$^\S$\footnote{email: jfc@posta.unizar.es}\footnote
{Corresponding author},
 Nathalie Debergh$^{\ddag}$\footnote{Chercheur, Institut Interuniversitaire des
Sciences Nucl\'eaires, Bruxelles, \\ email: Nathalie.Debergh@ulg.ac.be}
 and Giuseppe Marmo$^{\S\S}$\footnote{email: Giuseppe.Marmo@na.infn.it}}

\vskip 0.5cm

\centerline{$^{\ddag}$Theoretical and Mathematical Physics,Institute of Physics (B5),
 University of Li\`ege,}
\medskip
\centerline{ B-4000 Li\`ege 1 (Belgium)}

\bigskip

\centerline{$^\S$Departamento de  F\'{\i}sica Te\'orica, Universidad de Zaragoza,}
\medskip
\centerline{50009 Zaragoza, Spain.}

\bigskip
 
\centerline{$^{\S\S}$Dipartimento  di Scienze Fisiche, Universit\'a  Federico II
di Napoli and
INFN, Sezione di Napoli}
\medskip
\centerline{Complesso Universitario di Monte Sant'Angelo,
Via Cintia, 80125 Napoli, Italy}

\vskip 1cm

\begin{abstract}
Previous $\lambda$-deformed {\it non-Hermitian} Hamiltonians with respect to 
the usual scalar product of Hilbert spaces dealing with harmonic 
oscillator--like developments are (re)considered with respect to a 
new scalar product in order to take into account their property of 
self-adjointness. The corresponding deformed $\lambda$-states lead 
to new families of coherent states according to the DOCS, AOCS 
 and MUCS  points of view.
\end{abstract}

\newpage

Coherence and squeezing have recently been revisited \cite{{BDS98},{DBS}} 
through new families of states depending on an extra $\lambda$-parameter. 
This real parameter has been introduced by exploiting the {\it subnormal}
 character \cite{Sz}
 of the so-called (bosonic) {\it creation\/} operator acting on Fock spaces, 
a very well--known context inside harmonic oscillator--like developments.

A generalization of the factorization technique started by Schr\"odinger \cite{Sch}
was exploited in a recent paper \cite{DBS} for analyzing some previous results 
on non-hermitian oscillator-like Hamiltonians with real spectra.

We recall (see e.g. \cite{CMPR98} and references therein)
 that the classical factorization technique for a given Hamiltonian
of a one-dimensional quantum system 
\begin{equation}
H=-\frac 12\frac {d^2}{dx^2}+V(x)\label{Hamil}
\end{equation}
 consists in finding a constant $c$ and a 
superpotential function $W(x)$ such that
\begin{equation}
H-c=A^{\dag}A\ ,\label{factoriz}
\end{equation}
where
\begin{equation}
A=\frac 1{\sqrt 2}\left(\frac d{dx}+W(x)\right)  \ .\label{facA}
\end{equation}

In this case, the operator $A$ intertwines $H$ with the partner Hamiltonian
 $\wt H$ defined by $\wt H=AA^{\dag}+c$, i.e., $AH=\wt HA$. Moreover, the adjoint
operator $A^{\dag}$ intertwines $\wt H$ with $H$, i.e., $A^{\dag}\wt H=HA^{\dag}$.
In the particular case of the harmonic oscillator, 
\begin{equation}
H=-\frac 12\frac {d^2}{dx^2}+\frac {x^2}2\ ,\label{Hosc}
\end{equation} 
the well--known annihilation and creation operators 
\begin{equation}
a=\frac 1{\sqrt 2} \left(\frac d{dx}+x\right)\,,\qquad  a^{\dag}=
\frac 1{\sqrt 2} \left(-\frac d{dx}+x\right)\ ,\label{aosc}
\end{equation}
play the role of $A$ and $A^{\dag}$, while $c=\frac 12$. We recall that the operators $a$ and $a^{\dag}$, together with the identity $1$,
 generate the Heisenberg 
algebra characterized by 
\begin{equation}
[a,a^{\dag}]=1\ .\label{cra}
\end{equation}

As it is well known, the commutation relations $$[H,a]=-a\,,\qquad [H,a^{\dag}]=
a^{\dag}\ ,$$ can be used to obtain the full spectrum of the quantum harmonic 
oscillator operator. As $H>0$ there will be a state $\psi_0$ such that $a\psi_0=0$.
Once it has been normalized it is given by 
\begin{equation}
\psi_0 =\pi^{-1/4}e^{-\frac 12x^2}\ ,\label{gsosc}
\end{equation}
and the normalized eigenstates corresponding to the eigenvalues $E_n=(n+\frac 12)$,
 with $n=0,1,\ldots$, are
\begin{equation}
\psi_n(x)=\frac 1{\sqrt{n!}}\,(a^{\dag})^n\psi_0=  \frac{\pi^{-\frac{1}{4}}\, 2^{-\frac{n}{2}}}{\sqrt{n!}}\, e^{-\frac{1}{2}x^2} H_n(x)\ .\label{eigenosc}
\end{equation}
They form an orthonormal basis with respect to the standard inner product 
\be
\<\phi_1, \phi_2> = \int_{-\infty}^{+\infty} {\overline {\phi_1(x)}} \phi_2(x)\, dx\ ,
\label{sinprod}
\ee
 according to
\be
\int_{-\infty}^{+\infty} e^{-x^2}\, H_n(x)\,H_m(x) dx = \sqrt{\pi}\, 2^n\, n!\, \delta_{nm}\ ,
\ee
$H_n(x)$ being the very well-known Hermite polynomials \cite{MOS66}. In that context $x$ and $p=-i \frac{d}{dx}$ are self-adjoint operators associated with the important physical observables position and momentum, respectively.

Now, given a minimal numerable set of elements  $\{\varphi_0,\varphi_1,\ldots,\varphi_n,\ldots\}$
of a separable Hilbert  space $(\Hil,(\cdot,\cdot))$ whose closure is $\Hil$,
we can  introduce a new inner product in $\Hil$ by means of a bilinear form 
  $\<\\<\cdot,\cdot>>$ such that
$$\<\<\varphi_n,\varphi_m>>=\delta_{mn}\ ,
$$
and a pair of operators $A$ and $A^+$ such that
\begin{eqnarray}
&&A\varphi_0=0\ ,\\
&&A\varphi_n=
\sqrt n\,
\varphi_{n-1}\ ,\\
&&A^ {+}
\varphi_{n-1}=\sqrt n\,\varphi_n\ .
\end{eqnarray}

Consequently, the operators $A$ and $A^+$ so defined satisfy the 
Heisenberg algebra relation
\begin{equation}
[A,A^{+}]=1\ .\label{crA}
\end{equation}

Consider the Hamiltonian $H$ defined by 
\begin{equation}
H=\frac 12 \{A,A^+\}=\frac 12 (AA^++A^+A)=A^+A+\frac 12\ . \label{HA}
\end{equation}
If we use it to define equations of motion  we get the 
dynamics  of the harmonic oscillator. 

Indeed, as it was done in \cite{{AMM00},{MMP00}} for nonlinear coherent states,
we can define a map $D:{\Bbb C}\to {\rm Op\,}(\Hil)$ as follows:
$$D(z)=e^{(zA^{+}-\bar zA)}\ ,
$$
which defines a displacement operator. For any complex number $z$,
let us define  the state vectors $\ket {z}$ by $$\ket z=D(z)\varphi_0\ .
$$
This is but a Weyl system in the Bargmann--Fock representation. The
commutation rules of $A$, $A^+$ and $H$ being he same ones as for the 
standard one--dimensional harmonic oscillator, the dynamical evolution
 will preserve the set of such vectors  $\ket {z}$ and the
 dynamics associated with $H$ on $A$ and $A^+$  will define a dynamics on
 $\Bbb C$
by requiring it to be equivariant with respect to the $D$--map. Indeed, 
given a $z_0$ at $t=0$, the state evolves because $A$ and $A^+$ depends on $t$
and 
for any time $t$ we can define 
$$\ket{z_t}=e^{(z_tA^{+}-\bar z_tA)}\varphi_0\ ,
$$
giving rise, by derivation with respect to $t$ to
\begin{eqnarray}\frac {dz}{dt}&=i\, z\ ,\cr
\frac {d\bar z}{dt}&=-i\, \bar z\ .\nonumber
\end{eqnarray}

We can now consider the case of not only one, but a family $A^+_\lambda$
 of operators. The more general case of a two--parameter family proposed in \cite{WKO} is quite similar.
As already previously discussed \cite{BDS98}, let us introduce the creation operator in the following way
\be
A^+_\lambda=a^{\dagger}+\lambda\, I\,,\qquad  \lambda \in \
\R\,,\label{amasl}
\ee
where $a^{\dagger}$ is the adjoint of the annihilation operator $a$ 
given by (\ref{aosc}), together with $A_\lambda=a$, and define 
$$\phi_{n\,\lambda}=\frac 1{\sqrt{n!}}(A^+_\lambda)^n \psi_0\ ,
$$
where $\psi_0$ is the standard ground-state for the harmonic oscillator
given by (\ref{gsosc}). Notice that $\phi_{0\,\lambda}=\psi_0$ and $A^+_\lambda
\phi_{n\,\lambda}=\sqrt{n+1}\ \phi_{n+1\,\lambda}$.

As we suggested before we introduce a new 
inner product such that 
$$h_\lambda(\phi_{n\,\lambda},\phi_{m\,\lambda})=\<\langle\phi_{n\,\lambda},
\phi_{m\,\lambda}\rangle>=\delta_{nm}\ .
$$

We can now compute the adjoint of $A_\lambda^+$ with respect to the
inner product $\<\langle\cdot,\cdot\rangle>$. Indeed it is given by the relation 
$$ \<\langle\phi_{n\,\lambda},(A_\lambda^+)^{\dag} \phi_{m\,\lambda}\rangle>=\<\langle  A^+_\lambda
\phi_{n\,\lambda},
\phi_{m\,\lambda}\rangle>\ ,
$$
and therefore
$$\sqrt{n+1}\ \delta_{n+1,m}=\sqrt m\ \delta_{n,m-1}=\<\langle \phi_{n\,\lambda},
(A_\lambda^+)^{\dag}\phi_{m\,\lambda}\rangle>\ ,
$$
from which we see that the adjoint of $A^+_\lambda$ 
with respect  to $\<\langle\cdot,\cdot\rangle>$
is given by 
$$(A^+_\lambda)^{\dag}\phi _{n\,\lambda}=\sqrt n\,\phi_{n-1\,\lambda}\ ,$$
i.e.,   $(A^+_\lambda)^{\dag}=a$, for any value of the parameter $\lambda$.

As $a$ and $A^+_\lambda$ satisfy commutation relations as in (\ref{crA}),
we can consider a Hamiltonian like in (\ref{HA}), 
\begin{equation}
H_\lambda= A_\lambda^+ \, a+\frac 12=H_0+\lambda \,a\ ,
\end{equation}
and these developments lead to a {\it non-Hermitian} oscillator--like 
Hamiltonian \cite{BDS98} but with a {\it real} spectrum and specific 
eigenfunctions of special 
interest for discussing and comparing coherent as well as squeezed states 
with respect to already known ones \cite{{KS85},{Yu76}}. This is a direct 
consequence of the commutation rule $[A_\lambda,A^+_\lambda]=1$ and then
 $\widetilde H_\lambda=H_\lambda+1$. 

In fact, this non-Hermitian character was associated  with the
 scalar product (\ref{sinprod}) of the current Hilbert spaces we are dealing 
with in the 
conventional approaches. However, the Hamiltonian is hermitian with respect to
 the alternative inner product $\<\<\cdot,\cdot>>$, because the conjugate of $A_\lambda^+$
is $a$. The new product can be expressed by 
\begin{equation}
\<\<\ \varphi_1,\varphi_2>>= 
\int_{-\infty}^{+\infty} {\overline{ \varphi_1(x)}}\, \varphi_2(x)\rho(\lambda,
x)\, dx\ ,
\end{equation}
and the condition for $a$ to be the adjoint of  $A^{+}_{\lambda}$
 is 
\begin{equation}
\<\<\ \varphi_1, A^{+}_{\lambda}\varphi_2>>=\<\< a\varphi_1, \varphi_2 >>\ ,
\end{equation}
i.e.,
\begin{equation} 
\pd{\rho}{x}+{\sqrt 2} \, \lambda=0\ ,
\end{equation}
with general solution
\begin{equation}
\rho (\lambda , x) = C(\lambda)\, \exp\left(-{\sqrt{2}} \lambda x\right)\ .
\end{equation}

The arbitrary function $C(\lambda)$ can be fixed by imposing that $\<\<\phi_0,\phi_0>>=1$,
and we will obtain
\be
C(\lambda)=\exp\left(-\frac {\lambda^2}2\right)\ .
\ee

Mean values and information coming from Heisenberg relations are now easy to evaluate in the new $\lambda$-context. The following results come out by exploiting generalized Poisson integrals such as \cite{GH58}
\be
\int_{0}^{+\infty} e^{-l x^m} x^k dx = \frac{1}{m}\, l^{-\frac{k+1}{m}} \,
\Gamma\left(\frac{k+1}{m}\right)\,,\quad
l>0,\, m>0,\, k>-1\,.
\ee
We get
\be
\<x>_{\lambda} = -\frac{\lambda}{\sqrt{2}}\,,\quad  \<x^2>_{\lambda} =n+\frac{1}{2} +\frac{\lambda^2}{2}\,,\quad  (\Delta x)^2_{\lambda} = n+\frac{1}{2}\ ,
\ee
and
\be
\<p>_{\lambda} = -i\frac{\lambda}{\sqrt{2}}\,, \quad \<p^2>_{\lambda} =n+\frac{1}{2} -\frac{\lambda^2}{2}\,,\quad (\Delta p)^2_{\lambda} = n+\frac{1}{2}\ ,
\ee
showing that squeezing is not possible in this $\lambda$-context while coherence can be developed in the usual $n=0$-context \cite{{KS85},{Pe86}}. Moreover it is clear from these results that if $x$ is still Hermitian with respect to the new scalar product (18), $p$ is not. More precisely, it can be verified from (18) that the Hermitian conjugate of $p$ is $p + i\sqrt{2} \lambda$ implying that if we want to restore the Hermiticity for the momentum operator, we have to consider
\be
p_{\lambda}\equiv -i\frac{d}{dx} +i \frac{\lambda}{\sqrt{2}}\ ,
\ee
instead of $p$. We then have
\be
\<p_{\lambda}>_{\lambda} = 0\,,\quad \<p_{\lambda}^2>_{\lambda} =n+\frac{1}{2}\, ,\quad (\Delta p_{\lambda})^2_{\lambda} = n+\frac{1}{2}\,.
\ee
\par
Let us now exploit the previous results and turn to the construction of new coherent states depending on $\lambda$. Following the DOCS (displacement operator coherent spates) point of view \cite{Pe86}, it is straightforward to construct such states in terms of the new creation ($A^{+}_{\lambda}$) and annihilation ($a$) operators through the  displacement
 operator 
$$D_\lambda(z)=e^{(-\bar z a+z A_\lambda^{+}) }\ ,
$$
which allows us to introduce new coherent states,
$$\ket{z,\lambda}=D_\lambda(z)\psi_0\, ,
$$
whose coordinate representation will be 
\be
\psi_{z}(\lambda , x) = e^{z A^{\dagger}_{\lambda} - \bar z a} \psi_0(x) = e^{-\frac{1}{2}|z|^2} \sum_{n=0}^{\infty} \frac{z^n}{\sqrt{n!}} \phi_{n\lambda}(x)\ ,
\ee
which are such that
\be
a \,\psi_{z}(\lambda , x) = z\, \psi_{z}(\lambda , x)\ ,
\ee
ensuring also the AOCS (annihilation  operator coherent spates) point of view. 
Due to the fact that the normalization of these states with respect to the scalar
 product (18)
\be
\int_{-\infty}^{+\infty} {\overline{ \psi_{z}(\lambda , x)}}
\psi_{z}(\lambda , x) \rho (\lambda , x) dx = 1
\ee
is realized, we have the following mean values
\be
\<x>_{z} = \frac{1}{\sqrt{2}} (z + \bar z - \lambda)\ ,
\ee
\be
\<p_{\lambda}>_{z} = \frac{i}{\sqrt{2}} (-z + \bar z ) \ ,
\ee
and similar results concerning $<x^2>_{z}, <p^2_{\lambda}>_{z}$ finally leading to
\be
(\Delta x)^2_{z} = (\Delta p_{\lambda})^2_{z} = \frac{1}{2}\ .
\ee
The third and last point of view, the MUCS (minimal uncertainty coherent states) 
one, is thus also ensured in this context of new $\lambda$-coherent states.

Coming back to the classical interpretation,
 the configuration space of the system we are considering is the complex plane
 $\C$
that can be identified with ${\R}^2$ according to 
 
\begin{equation}
z=\frac 1{\sqrt 2}(q+i\,p)\,, \qquad \bar z=\frac 1{\sqrt 2}(q-i\,p)\,. \label{CR2id}
\end{equation}
In this way we are  choosing one of the
possible  complex structures in $\R^2$.
We can define the inner product in $\C$ given by 
 $h(z,w) = \bar{z}w$, whose real part
 provides one half of the standard  Euclidean structure in ${\R}^ 2$, while the 
imaginary part gives us  one half of 
the standard  symplectic structure. The complex 
structure in   ${\R}^ 2$ is  
 defined by
multiplication by $i$, i.e., $Jz = i\,z$, then $J(q,p) = (-p,q)$.
 The tangent space to  ${\Bbb
R}^2$ at each point $m\in {\Bbb R}^2$ is identified in a natural way to
${\Bbb R}^2$ itself as a linear space. A local basis of the module  of
 vector fields in ${\Bbb R}^2$ is made up by 
$\partial/\partial q$ and $\partial/\partial p$,
and that of the differential forms  
by $ dq$ and $dp$. 
The complexified of this space is also generated by 
\begin{equation}
 dz = \frac 1{\sqrt 2}(dq + i\, dp)\,, \qquad 
d\bar  z = \frac 1{\sqrt 2}(dq -
i\, dp)\, ,
\end{equation} 
and  similarly, for the dual basis of vector fields,
\begin{equation}
\pd{}z = \frac 1{\sqrt 2} \left( \pd{}q - i\pd
{}p \right)\ , ~~~~~ \pd{}{\bar z} = \frac 1{\sqrt 2} \left ( \pd {}q + i\pd
{}p \right)\ .
\end{equation}

In these complex coordinates, the canonical symplectic form is 
$$\omega=i\, dz\wedge d\bar z\ .
$$

Now, when comparing (\ref{aosc}) with  (\ref{CR2id}) we see that 
(\ref{amasl}) can be seen in the classical approach as a transformation 
$$z\mapsto z_\lambda=z\,,\qquad \bar z\mapsto \bar z_\lambda=\bar z+\lambda\ ,
$$
which is a canonical transformation, but it does not preserve the Hamiltonian
for the harmonic oscillator $H(z,\bar z)=h(z,\bar z)$,
which transforms into 
\begin{equation}
H_\lambda=H+\lambda \,z\ .
\end{equation}

Therefore the dynamics 
$$\Gamma =i\left(z\pd{}z-\bar z\pd{}{\bar z}\right)\ ,
$$
changes to
$$\Gamma_\lambda=\Gamma-i\,\lambda \pd{}{\bar z}\ .$$

 However, the transformation being canonical, the 
image of closed orbits for the harmonic oscillator still remain closed, and 
moreover, the period is also constant as for the harmonic oscillator, therefore 
the quantum spectrum is equally spaced.  

In summary, in the new variables the system is indeed an Harmonic
Oscillator (both at the classical level and at the quantum level). However
from the physical point of view, when we compare them in the same
coordinate system, we find that they identify different physical systems
because the equilibria points are different (at the classical level) and
the vacua states are different at the quantum level. 
Even though they are represented {\sl abstractly\/} by the algebra of
the Harmonic Oscillator, their {\sl realizations\/} identify different {\sl physical
oscillators\/} characterized by different zero modes.
Thus the dynamics in the {\sl lambda variables\/} has the same form as the other
dynamics in the original coordinates. To compare them however, we need to
express both in the same coordinate system and when this is done they are
associated with different vector fields.


\begin{thebibliography}{AAAA 1999}

\bibitem
{BDS98}
J. Beckers, N. Debergh and F.H. Szafraniec, 
Phys. Lett. {\bf A 243} (1998) 256--60,
Phys. Lett. {\bf A 246} (1998) 561--561.


\bibitem
{DBS}
N. Debergh, J. Beckers  and F.H. Szafraniec, 
Phys. Lett. {\bf A 267} (2000) 113-16.


\bibitem
{Sz} F.H. Szafraniec,
 Comm. Math. Phys. {\bf 210} (2000) 323--34 and references therein.


\bibitem
{Sch} E. Schr\"odinger E.,
Proc. Roy. Irish Acad. {\bf A 46} (1940) 9--16;
 Proc. Roy. Irish Acad. {\bf A 46} (1940) 183--206;
Proc. Roy. Irish Acad. {\bf A 47} (1941) 53--54.

\bibitem
{CMPR98}
J. F. Cari\~nena,  G. Marmo,  A. M. Perelomov and  M. F. Ra\~nada:
 Int. J. Mod. Phys. {\bf A 13}  (1998) 4913--29.


\bibitem
{MOS66} W. Magnus, F. Oberhettinger, R.P. Soni, 
Formulas and Theorems for the Special Functions of Mathematical Physics 
(Springer, Berlin, 3rd Edition, 1966).

\bibitem
{AMM00}
P. Aniello, V. Man'ko, G. Marmo, S. Solimeno and F. Zaccaria, 
 J.  Opt. B:Quantum Semiclass. Opt.  {\bf 2}, (2000) 718--25.

\bibitem
{MMP00}
 V. Man'ko, G. Marmo, A. Porzio, S. Solimeno and F. Zaccaria, 
 Phys. Rev. {\bf A 62}, 053407 (2000).

\bibitem%
{WKO}X. Wang, L.C. Kwek and C.H. Oh, Phys. Lett. {\bf A 259} (1999) 7-14.

\bibitem
{KS85} J.R. Klauder and B.S. Skagerstam, Coherent States, Applications in Physics and Mathematical Physics (World Scientific, Singapore, 1985).

\bibitem
{Yu76} H.P. Yuen, Phys. Rev. {\bf A 13} (1976) 2226.

\bibitem
{GH58} W. Grobner  and N. Hofreiter, Integraltafel, Zweiter Teil, Bestimmte 
Integrale (Springer, Wien, 1958).

\bibitem
{Pe86} A.M. Perelomov, Generalized Coherent States and their Applications (Springer, New York, 1986).

\end{thebibliography}
\end{document}